\documentclass[12pt,a4paper]{article}
\usepackage{amssymb}
\newcommand{\initiate}{\setcounter{equation}{0}}
\newcounter{eg}                                 
\newtheorem{eg}{Example}[section]               
\def\beg{\begin{eg}\rm}                         
\def\eeg{\hfill\sq\end{eg}}                     
\def\c#1{{\cal #1}}                             

\def\Dirac{{\raise0.09em\hbox{/}}\kern-0.69em D}
\def\h#1{\hat{#1}}                              
\def\kbar{{\mathchar'26\mkern-9muk}}            
\def\p{\partial}                                
\def\sq{\hbox{\rlap{$\sqcap$}$\sqcup$}}         
\def\t#1{\tilde{#1}}                            
\def\tfrac #1#2{\textstyle{\frac{#1}{#2}}}      

\def\k{\kern-.1em\mathbin{,}\kern-.1em}
\def\hk{\kern.12em\raise-1em\hbox{$\hat{\raise1em\hbox{,}}$}\kern.12em}
\def\be{\begin{equation}}                     
\def\bea{\begin{eqnarray}}                      %
\def\ee{\end{equation}}                         %
\def\eea{\end{eqnarray}}                        %
\begin{document}

\title{Gravity and the Structure of\\
Noncommutative Algebras}

\author{M. Buri\'c$\strut^1$, \ \ T. Grammatikopoulos$\strut^2$, \\
        J. Madore$\strut^3$, \ \ G. Zoupanos$\strut^2$ \\[30pt]
      $\strut^1$Faculty of Physics, P.O. Box  368, 11001 Belgrade\\
\and  $\strut^2$Physics Department, National Technical University\\
       Zografou Campus, GR-157~80 Zografou, Athens
\and  $\strut^3$Laboratoire de Physique Th\'eorique\\
      Universit\'e de Paris-Sud, B\^atiment 211, F-91405 Orsay}

\date{}

\maketitle

\abstract {A gravitational field can be defined in terms of a moving
  frame, which when made noncommutative yields a preferred basis for a
  differential calculus. It is conjectured that to a linear perturbation of
  the commutation relations which define the algebra there corresponds
  a linear perturbation of the gravitational field. This is shown to
  be true in the case of a perturbation of Minkowski space-time.}

\parskip 10pt plus2pt minus2pt
\parindent 0pt
\vfill
\medskip
\thispagestyle{empty} \newpage


\setlength{\baselineskip}{15pt}
\setlength{\parskip}{15pt plus2pt minus2pt}

\initiate
\section{Introduction}

It is known that to a noncommutative geometry one can associate in
various ways a gravitational field. This can be elegantly
done~\cite{Con88,Con96} in the imaginary-time formalism and perhaps
less so~\cite{MadMou98} in the real-time formalism. We examine here
the inverse problem, that of associating a noncommutative geometry to
a given classical field. As concrete examples, one would like to know
to what extent it is possible to give a noncommutative extension of
the Schwarzschild metric or of a cosmological metric. We would also
like to know how many extensions there are and what their properties.
There have been examples
constructed~\cite{MacMadManZou03,BurGraMadZou06b,BurMad05b}
more-or-less {\it ad hoc}; we give here a more systematic analysis by
restricting our considerations to the `semi-classical' theory,
retaining only contributions of first-order in the noncommutativity
parameter. As a working hypothesis we shall suppose that there is one
physical property, which at large scales manifests itself as gravity
and at small scales as noncommutativity.

When a noncommutative geometry is considered to lowest order the
commutativity relations define a symplectic form; the metric defines
also a curvature. We examine the relation between these two structures
which are imposed by the requirements of noncommutative geometry.  We
show that in certain simple situations the field `almost' determines
the structure of the algebra as well as the differential calculus.
Over a given algebra there can be many differential calculi but all
must satisfy certain consistency conditions before they can be
considered as associated to the algebra.  We are therefore interested
in the cases where these relations determine the field. We shall
consider almost exclusively the almost-commutative limit.  We shall
also briefly consider a sort of modified form of the background-field
approximation in which we suppose that to the structure of a
noncommutative algebra and associated differential calculus there has
been associated a geometric structure and we proceed so to speak by
induction to extend the correspondence to a first-order perturbation.

As a measure of noncommmutativity, and to recall the many parallelisms
with quantum mechanics, we use the symbol $\kbar$, which will
designate the square of a real number whose value could lie somewhere
between the Planck length and the proton radius $m_P^{-1}$. Although
this is never explicitly used we shall think rather of the former and
identify $\kbar$ with Newton's constant $G_N$ (in units with
$\hbar=1$). This becomes important when we consider perturbations. 
We introduce a set $J^{\mu\nu}$ of elements of an associative algebra $\c{A}$
and use them to define commutation relations
\begin{equation}
[x^\mu, x^\nu] = i\kbar J^{\mu\nu}(x^\sigma).        \label{xx}
\end{equation}
The $J^{\mu\nu}$ are of course restricted by Jacobi identities; we see
below that there are two other natural requirements which also
restrict them.

Let $\mu$ be a typical `large' source mass with `Schwarzschild radius'
$G_N\mu$. If noncommutativity is not directly related to gravity then
it makes sense to speak of ordinary gravity as the limit $\kbar \to 0$
with $G_N\mu$ non vanishing. On the other hand if noncommutativity and
gravity are directly related then both should vanish with $\kbar$.
The two points of view are not at odds provided one considers
(classical) gravity as a purely macroscopic phenomenon, valid only for
`large' masses.  We shall use the dimensionless parameter
\begin{equation}
\epsilon = \kbar \mu^2                           \label{ekm}
\end{equation}
as a measure of the relative importance of noncommutative effects.  
We shall also use the WKB formalism to illustrate the close relation which
exists between $J^{\mu\nu}$ and the geometry of the wave.  The WKB
approximation is a classical description of quantum mechanics in the
sense that derivations can be identified with the momenta.  We
shall show that in the presence of the wave Jacobi identities require a
modification of the structure of the algebra.  Let $\omega$ be the
characteristic mass scale associated with the wave. We shall require the
inequalities
\begin{equation}
\sqrt \kbar \ll \omega^{-1} \ll \mu^{-1}.          \label{oom}
\end{equation}
If we consider $\kbar$ to be of the order of the Planck mass then the
first inequality states that the Planck mass is `large'; the second is
the definition of what is meant by a `high-frequency' wave.

The extra momenta $p_\alpha$ which must be added to the algebra in
order that the derivations be inner stand in duality to the position
operators $x^\mu$ by the relation
\begin{equation}
[p_\alpha, x^\mu] = e^\mu_\alpha.                 \label{pxe}
\end{equation}
The right-hand side of this identity defines the gravitational
field. The left-hand side must obey Jacobi identities. These
identities yield relations between quantum mechanics in the given
curved space-time and the noncommutative structure of the
algebra. The three aspects of reality then, the curvature of
space-time, quantum mechanics and the noncommutative structure of
space-time are intimately connected. We shall consider here the
even more exotic possibility that the field equations of general
relativity are encoded also in the structure of the algebra so that
the relation between general relativity and quantum mechanics can be
understood by the relation which each of these theories has with
noncommutative geometry.

In spite of the rather lengthy formalism the basic idea is simple. We
start with a classical geometry described by a moving frame
$\theta^\alpha$ and we associate 
\begin{equation}
 \theta^\alpha \buildrel \rho \over \longrightarrow J^{\mu\nu} \label{map}
\end{equation}
to it a noncommutative algebra with generators $x^\mu$ and commutation
relations~(\ref{xx}) which we identify with position space. To this
algebra we add the extra elements which are necessary in order that
the derivations become inner; this is ordinary quantum mechanics. The
new element is the fact that if the original algebra describes a
curved space-time then Jacobi identities force the extended algebra to
be noncommutative. More details of the map~(\ref{map}) will be given
in Section~\ref{agm}.

Typically one would proceed in three steps.  First choose a moving
frame to describe a metric. Quantize it by replacing the moving frame
by a frame, as described below. The important special cases referred
to above would include those frames which could be quantized without
ordering problems. Finally one looks for a noncommutative algebra
consistent with the resulting differential calculus; this is the
image of the map~(\ref{map}).  Let $e_\alpha$ be dual to the left-hand side
of~(\ref{map}).  If we quantize as in~(\ref{pxe}) by imposing the rule
\begin{equation}
e_\alpha \mapsto p_\alpha
\end{equation}
then from the Jacobi identities we find
\begin{equation}
[p_\alpha, J^{\mu\nu}] = [x^{[\mu},[p_\alpha,x^{\nu]}]].
\end{equation}
If the space is flat and the frame is the canonical flat frame then
the right-hand side vanishes and it is possible to consistently choose
the expression $J^{\mu\nu}$ to be equal to a constant or even to
vanish.  But on the other hand, if the space is curved the right-hand
side cannot vanish identically; we must conclude then that
$J^{\mu\nu}$ is non-trivial. This means that the kernel of the
map~(\ref{map}) must vanish. On the other hand it cannot be single
valued for any constant $J$ has flat space as inverse image.

The physical idea we have in mind has been given
elsewhere~\cite{Mad89c,Mad00b,Mad00c}. One can use a solid-state
analogy and think of the ordinary Minkowski coordinates as macroscopic
order parameters obtained by `coarse-graining' over regions whose size
is determined by a fundamental area scale $\kbar$, which is
presumably, but not necessarily, of the order of the Planck area
$G\hbar$. They break down and must be replaced by elements of a
noncommutative algebra when one considers phenomena on smaller scales.
A simple visualization is afforded by the orientation order parameter
of nematic liquid crystals. The commutative free energy is singular in
the core region of a disclination. There is of course no physical
singularity; the core region can simply not be studied using the
commutative order parameter.

There is also a certain similarity with the effect of screening in
quantum field theory and in plasma physics. One can consider a `point'
as surrounded by a `cloud of void' which `screens' it from
neighbouring `points'. Because the commutator defines in the
commutative limit an antisymmetric tensor field there are obvious
analogies with spin and with the electromagnetic field; we have not
however found any particularly fruitful insights using these.  
In the limit when the gravitational field vanishes there still remains
a definite frame at each point defined, for example, by the Petrov
vectors. So even in flat space, if considered as the result of such a
limiting process, local Lorentz invariance is broken. This residual
memory could be considered to be similar to that invoked in Mach's
principle.  Other reasons have been proposed~\cite{Dop01} for this breaking.

A detailed description of the method we shall use has been given in a
previous article~\cite{MacMad03} and it suffices therefore here to
outline the prescription.  We suppose that a complete consistent
noncommutative geometry has been given. By this we mean that the frame
and the commutation relations are explicitly known. We shall perturb
both the geometry and the algebra and show that the perturbation of
the one can be intimately related with that of the other such that the
resulting geometry is consistent.  The fact that the geometry depends
only on the formal algebraic structure of the algebra, seemingly
independent of the representation is perhaps due to the fact that only
first-order perturbations are explicitly calculated.  Although one
cannot claim to have defined completely an algebra without a choice of
state, we have not found it necessary to use a concrete representation
in the calculations we have presented here. This is certainly related
to the fact that most concrete calculations are presented only in the
quasi-classical approximation.  Although noncommutative `gravity' in
the Kaluza-Klein sense had been investigated
earlier~\cite{Mad89c,MadMou93a} it would seem that the first concrete
example of noncommutative `gravity' was~\cite{Mad92a} an extension of
the 2-sphere. Although not very interesting as a realistic example of
gravity it clearly illustrates the relation between the commutation
relations and the effective classical gravitational field.  There have
been several recent investigations of the same subject, at least two of
which~\cite{AscDimMeyWes05,LanSza01} are not far in spirit from the
present calculations.

\initiate
\section{General considerations}

Let then $\c{A}$ be a noncommutative $*$-algebra generated by four
hermitian elements $x^\mu$ which satisfy the commutation
relations~(\ref{xx}).  Assume that over $\c{A}$ is a differential
calculus which is such~\cite{Mad00c} that the module of 1-forms is free and
possesses a preferred frame $\theta^\alpha$
which commutes, 
\begin{equation}
[x^\mu, \theta^\alpha] = 0,                             \label{mod}
\end{equation}
with the algebra. The space one obtains in the commutative limit is
therefore parallelizable with a global moving frame
$\t{\theta}^\alpha$ defined to be the commutative limit of
$\theta^\alpha$.  We can write the differential
\begin{equation}
dx^\mu = e_\alpha^\mu \theta^\alpha, \qquad e_\alpha^\mu = e_\alpha x^\mu.
\end{equation}
The algebra is defined by a product which is restricted by the matrix
of elements $J^{\mu\nu}$; the metric is defined, we shall see below, by
the matrix of elements $e_\alpha^\mu$. Consistency requirements,
essentially determined by Leibniz rules, impose relations between these
two matrices which in simple situations allow us to find a one-to-one
correspondence between the structure of the algebra and the metric.
The input of which we shall make the most use is the Leibniz rule
\begin{equation}
i\kbar e_\alpha J^{\mu\nu} =
[e^\mu_\alpha, x^\nu] - [e^\nu_\alpha, x^\mu].      \label{lr}
\end{equation}
One can see here a differential equation for $J^{\mu\nu}$ in terms of
$e^\mu_\alpha$. In important special cases the equation reduces to a
simple differential equation of one variable. 

The relation~(\ref{lr}) can be written also as Jacobi identities 
\begin{equation}
[p_\alpha, [x^\mu, x^\nu]] +
[x^\nu, [p_\alpha, x^\mu]] +
[x^\mu, [x^\nu, p_\alpha]] = 0
\end{equation}
if one introduce the momenta $p_\alpha$ associated to the derivation
by the relation~(\ref{pxe}).

Finally, we must insure that the differential is well defined. A
necessary condition is that $d[x^\mu, \theta^\alpha] = 0$.  It follows
that
\begin{equation}
d[x^{\mu}, \theta^{\alpha}] =
[dx^{\mu}, \theta^{\alpha}] + [x^\mu, d\theta^{\alpha}] =
e^\mu_\beta [\theta^{\beta}, \theta^{\alpha}] -
\tfrac 12 [x^\mu, C^\alpha{}_{\beta\gamma}] \theta^{\beta}\theta^{\gamma}.
\end{equation}
We have here introduced the Ricci rotation coefficients 
$C^\alpha{}_{\beta\gamma}$.
We find then that multiplication of 1-forms must satisfy
\begin{equation}
[\theta^{\alpha}, \theta^{\beta}] =
\tfrac 12 \theta^\beta_\mu [x^\mu, C^\alpha{}_{\gamma\delta}]
\theta^{\gamma}\theta^{\delta}.                               \label{bffp}
\end{equation}
Consistency requires then that
\begin{equation}
\theta^{[\beta}_\mu [x^\mu, C^{\alpha]}{}_{\gamma\delta}] = 0. \label{xC}
\end{equation}
We have in general three consistency equations which must be
satisfied in order to obtain a noncommutative extension. They are the
Leibniz rule~(\ref{lr}), the Jacobi identity and the condition
(\ref{xC}) on the differential.  The first two constraints follow from
Leibniz rules but they are not completely independent of the
differential calculus since one involves the momentum operators.

To illustrate the importance of the Jacobi identities we mention that
they force a modification of the canonical commutation relations and
introduce a dependence 
\begin{equation}
\hbar\delta^\mu_\alpha \mapsto  \hbar e^\mu_\alpha
\end{equation}
of Planck's `constant' on the gravitational field. We mentioned
already that if one place the canonical commutator~(\ref{pxe}) in the
Jacobi identity with two coordinate and one momentum entry that for
this to be consistent the coordinates in general cannot commute.

\initiate
\section{Linear perturbations of flat space}   \label{gf}

If we consider the $J^{\mu\nu}$ of the previous sections as the
components of a classical field on a curved manifold then in the limit
when the manifold becomes flat the `equations of motion' are Lorentz
invariant. We notice however that in this limit they are also
degenerate. In particular solutions of the form~(\ref{JNe}) are
unacceptable. To remedy this we suppose that 
as $e^\lambda_\alpha \to e^\lambda_{0\alpha}$ we obtain
\begin{equation}
J^{\mu\nu} \to J_0^{\mu\nu}, \qquad \det J_0 \neq 0.
\end{equation}
Were we to choose $e^\lambda_{0\alpha}$ to be a flat frame then the
assumption would mean that $J_0^{\mu\nu}$ `spontaneously' breaks Lorentz
invariance. Since Lorentz invariance is broken for every non-flat
frame by definition, it would be a stronger assumption to suppose that
$J_0^{\mu\nu} = 0$.

We shall now consider fluctuations around a particular given solution
to the problem we have set. We suppose that is we have a reference
solution comprising a frame $e^\lambda_{0\alpha} = \delta^\lambda_{\alpha}$ 
and a commutation relation $J_0^{\mu\nu}$ which we perturb to
\begin{equation}
J^{\alpha\beta} = J_0^{\alpha\beta} + \epsilon I^{\alpha\beta}, \qquad 
e^\mu_\alpha = \delta^\mu_{\beta}
(\delta^\beta_\alpha + \epsilon\Lambda^\beta_\alpha).     \label{pert}
\end{equation}
In terms of the unknowns $I$ and $\Lambda$ the Jacobi and Leibniz
constraints become respectively
\begin{eqnarray}
&&
\epsilon_{\lambda\mu\nu\sigma} [x^\lambda, I^{\mu\nu}] = 0,    \label{*1a}
\\[6pt]&&
e_\alpha I^{\mu\nu} = 
[\Lambda^\mu_\alpha, x^\nu] -  [\Lambda^\nu_\alpha, x^\mu].    \label{*2a}
\end{eqnarray}
We now use the fact, well known from quantum mechanics, that when the
value of the commutator is a constant then the commutator is a
derivative. That is, for any $f$
\begin{equation}
[x^\lambda, f] = i\kbar J_0^{\lambda\sigma} \p_\sigma f + o(\epsilon^2), \qquad
[p_\alpha, f] = \p_\alpha f + o(\epsilon^2).
\end{equation}
The two constraint equations become
\begin{eqnarray}
&&
\epsilon_{\lambda\mu\nu\sigma} 
J^{\lambda\sigma} \p_\sigma I^{\mu\nu} = 0,                  \label{*1b}
\\[6pt]&&
e_\alpha I^{\mu\nu} = 
\p_\sigma \Lambda^{[\mu}_\alpha J_0^{\sigma\nu]}.             \label{*2b}
\end{eqnarray}
These two equations are the origin of the particularities of our
construction, they and the fact that the `ground-state' value of
$J^{\mu\nu}$ is an invertible matrix.

The constraint equations become particularly transparent if one
introduce the new unknowns
\begin{equation}
\h{I}_{\alpha\beta} = 
J^{-1}_{0\alpha\gamma} J^{-1}_{0\beta\delta} I^{\gamma\delta}, \qquad
\h{\Lambda}_{\alpha\beta} = J^{-1}_{0\beta\gamma}\Lambda^\gamma_\alpha. \label{Hat}
\end{equation}
We decompose also $\h{\Lambda}$ as the sum 
\begin{equation}
\h{\Lambda}_{\alpha\beta} = 
\h{\Lambda}^+_{\alpha\beta} + \h{\Lambda}^-_{\alpha\beta}
\end{equation}
of a symmetric and antisymmetric term. The constraints become
\begin{eqnarray}
&&
e_\alpha (\h{I} + \h{\Lambda}^-)_{\beta\gamma} +
(e_\alpha\h{\Lambda}^-_{\beta\gamma} +
e_\beta\h{\Lambda}^-_{\gamma\alpha} +
e_\gamma\h{\Lambda}^-_{\alpha\beta}) =
e_{[\beta} \h{\Lambda}^+_{\gamma]\alpha},                        \label{L1b}
\\[6pt] &&
\epsilon^{\alpha\beta\gamma\delta}
e_\alpha (\h{I} + 2 \h{\Lambda})_{\beta\gamma} = 0.            \label{L2b}
\end{eqnarray}
We introduce
\begin{equation}
\h{I} = \tfrac 12 \h{I}_{\alpha\beta}\theta^\alpha \theta^\beta, \qquad
\h{\Lambda}^- = \tfrac 12 \h{\Lambda}^-_{\alpha\beta}\theta^\alpha \theta^\beta.
\end{equation}
The constraints simplify to `cocycle' conditions.  If we
multiply~(\ref{L1b}) by $\epsilon^{\alpha\beta\gamma\delta}$ we obtain
\begin{equation}
\epsilon^{\alpha\beta\gamma\delta} 
e_\alpha (\h{I} + 4\h{\Lambda}^-)_{\beta\gamma} = 0.          \label{+2}
\end{equation}
It follows then that
\begin{equation}
d\h{\Lambda}^- = 0, \qquad d\h{I} = 0.                  \label{dI}
\end{equation}
We can rewrite~(\ref{L1b}) as
\begin{equation}
e_\alpha (\h{I} + \h{\Lambda}^-)_{\beta\gamma} = 
e_{[\beta} \h{\Lambda}^+_{\gamma]\alpha}.                   \label{+3}
\end{equation}
This equation has the integrability conditions
\begin{equation}
e_\alpha e_{[\beta} \h{\Lambda}^+_{\gamma]\delta} -
e_\delta e_{[\beta} \h{\Lambda}^+_{\gamma]\alpha} = 0.            
\end{equation}
But the left-hand side is the linearized approximation to the
curvature of a metric with components 
$g_{\mu\nu} + \epsilon\h{\Lambda}^+_{\mu\nu}$. If it vanishes then the
perturbation is a derivative; for some 1-form $A$
\begin{equation}
\h{\Lambda}^+_{\beta\gamma} = \tfrac 12 e_{(\beta} A_{\gamma)}.   \label{dL+}
\end{equation}
Equation~(\ref{+3}) becomes therefore
\begin{equation}
e_\alpha (\h{I} + \h{\Lambda}^- - dA)_{\beta\gamma} = 0.    \label{+}
\end{equation}
It follows then that for some 2-form $c$ with constant components
$c_{\beta\gamma}$
\begin{equation}
\h{\Lambda}^- = - \h{I} + dA + c.                         \label{LI}
\end{equation}
The remaining constraints are satisfied identically. 
The most important relation is Equation~(\ref{LI}) which,
in terms of the original `unhatted' quantities, becomes
\begin{equation}
\Lambda^\alpha_\beta = 
J^{-1}_{0\beta\gamma} I^{\alpha\gamma} +
J_0^{\alpha\gamma}(c_{\beta\gamma} + e_\beta A_\gamma).         \label{wh}
\end{equation}
This condition is much weaker than, but similar to Equation~(\ref{pJx}).

\initiate
\section{The algebra to geometry map}                   \label{agm}

We can now be more precise about the map~(\ref{map}).  Let
$\theta^\alpha$ be a frame which is a small perturbation of a flat
frame and let $J^{\alpha\beta}$ be the frame components of a small
perturbation of a constant `background' $J_0$. Us interests the 
map
\begin{equation}
I^{\alpha\beta}  \buildrel \sigma \over
\longrightarrow \Lambda^\alpha_\beta 
= J^{-1}_{0\beta\gamma} I^{\alpha\gamma} +
J_0^{\alpha\gamma}(c_{\beta\gamma} + e_\beta A_\gamma).     \label{**}
\end{equation}
We recall that we are considering only first-order fluctuations around
a given frame and that these fluctuations are redundently
parameterized by the array $\Lambda^\alpha_\beta$. We can rewrite the
map $\rho$ as a map
\begin{equation}
\Lambda^\alpha_\beta  \buildrel \rho \over
\longrightarrow I^{\alpha\beta}.
\end{equation}
It can be defined as an inverse of the map $\sigma$ defined in
Equation~(\ref{**}). 

If we neglect all terms which are gradients then
we see that the extension of $\sigma$ to the metric is given by
\begin{equation}
g^{\alpha\beta} =  g_0^{\alpha\beta} - \epsilon h^{\alpha\beta}, \qquad
h^{\alpha\beta} = \Lambda^{(\alpha\beta)} = 
J_0^{(\alpha\gamma} \h{I}_{\gamma}{}^{\beta)}.
\end{equation}
We recall that a perturbation of a frame
\begin{equation}
e^\mu_\alpha = e^\mu_{0\beta}
(\delta^\beta_\alpha + \epsilon\Lambda^\beta_\alpha)     \label{pert0}
\end{equation}
engenders a perturbation
\begin{equation}
g^{\mu\nu} =  g_0^{\mu\nu} - \epsilon h^{\mu\nu}, \qquad
h^{\mu\nu} = \Lambda^{(\mu\nu)}
\end{equation}
of the metric.

There is a certain ambiguity in the map $\sigma$ defined
in~(\ref{**}). This must be so since over any associative algebra
there are many differential calculi. As an example of this one can
consider the case of constant commutators. The two key formulae are
\begin{equation}
\theta^\alpha = \theta_0^\alpha - 
\epsilon \Lambda^\alpha_\beta \theta_0^\beta, \qquad 
J^{\mu\nu} = J_0^{\mu\nu}
\end{equation}
The momenta are linear functions of the position and there is but one
calculus based on the derivations associated to the momenta. It is
given by the duality relations
\begin{equation}
dx^\mu (e_\alpha) = [p_\alpha, x^\mu] = \delta^\mu_\alpha.
\end{equation}
So amongst the set of differential calculi there is one which is based
on the derivations defined by the momenta. This is the one which we
define to be the image of $\sigma$.

Suppose we were to chose another `nearby', based on the frame 
\begin{equation}
\theta^\alpha = dx^\alpha - \epsilon \Lambda^\alpha_\beta dx^\beta
\end{equation}
and defined by some matrix $\Lambda^\alpha_\beta$.  We use the fact
that the formulae of Section~\ref{gf} remain valid but with the extra
condition that $I=0$. In particular, from Equations~(\ref{LI}) we find
that the 2-form $\h{\Lambda}$ is a coboundary. It does not contribute
to the Riemann tensor. So the perturbed differential calculus
engenders a trivial perturbation of the metric.  This result is
difficult to understand intuitively since one would expect the metric
components to change if the symmetric part $\Lambda_{(\alpha\beta)}$
of $\Lambda_{\alpha\beta}$ does not vanish.  However the Jacobi
identities force $\Lambda_{(\alpha\beta)}$ to be the symmetric
gradient of a 1-form $A$; it therefore does not contribute to the
Riemann tensor.

\initiate 
\section{Phase space}

It is obviously the case that in the commutative limit the 4
coordinate generators tend to the space-time coordinates and the 4
momenta tend to the conjugate momenta. The 8 generators become the
coordinates of phase space. For this to be consistent all Jacobi
identities must be satisfied, including those with two and three
momenta. We consider first the identities
\begin{equation}
[p_\alpha, [p_\beta, x^\mu] ] +  
[p_\beta, [x^\mu, p_\alpha] ] +                     
[x^\mu, [p_\alpha, p_\beta] ] = 0.                  
\end{equation}
One easily see that, using the identities~(\ref{dI}) and~(\ref{dL+}) as
well as the assumption that the center is trivial we find that
\begin{equation}
i\kbar [p_\alpha, p_\beta] = (K - \epsilon (\h{\Lambda} - dA))_{\alpha\beta}
= (K +\epsilon \h{I})_{\alpha\beta}
\end{equation}
with 
\begin{equation}
K = - J_0^{-1}.
\end{equation}
That is, 
\begin{equation}
i\kbar [p_\alpha, p_\beta] = - J^{-1}_{\alpha\beta} + o(\epsilon^2).
\end{equation}
The remaining identities, involving only the momenta, are then
satisfied by virtue of the fact that the 2-form $\h{\Lambda}$ is
closed. There is evidence to the fact that this relation is valid to
all orders in $\epsilon$.

From the Jacobi identities we find that
\begin{equation}
[p_\alpha - J^{-1}_{0\alpha\mu} x^\mu, x^\nu] = 
\delta^\nu_\alpha - J^{-1}_{0\alpha\mu} (J_0^{\mu\nu} + \epsilon I^{\mu\nu})
+ \epsilon \Lambda^\nu_\alpha
= \epsilon (\Lambda^\nu_\alpha - J^{-1}_{0\alpha\mu}  I^{\mu\nu} ) = 0.
\end{equation}
For some set of constants $c_\alpha$ therefore, if the center of the
algebra is trivial, we can write
\begin{equation}
i\kbar p_\alpha = J^{-1}_{0\alpha\mu} x^\mu + c_\alpha.           \label{pJx}
\end{equation}
The `Fourier transform' is linear. 

Let $J_0^{\mu\alpha}$ be an invertible matrix of real numbers. For each
such matrix there is an obvious map from the algebra to the geometry
given by
\begin{equation}
J^{\mu\nu}{} \mapsto
e^\nu_\alpha  =  J^{-1}_{0\alpha\mu} J^{\mu\nu}.              \label{JNe}
\end{equation}
For such frames we introduce momenta $p_\alpha$ and
find that
\begin{equation}
[p_\alpha, x^\nu] = e^\nu_\alpha = J^{-1}_{0\alpha\mu} J^{\mu\nu} = 
(i\kbar)^{-1} J^{-1}_{0\alpha\mu} [x^\mu, x^\nu].
\end{equation}
That is
\begin{equation}
[i\kbar p_\alpha - J^{-1}_{0\alpha\mu} x^\mu, x^\nu] = 0.               \label{px}
\end{equation}
We can conclude therefore that~(\ref{pJx}) is satisfied.  We can
interpret the results of the previous section as the statement that
this condition is stable under small perturbations of the geometry or
algebra.

\initiate 
\section{An Example}

Consider ($2-d$)-Minkowski space with coordinates $(t,x)$ which
satisfy the commutation relations $[t,x] = ht$ and with a geometry
encoded in the frame $\theta^1= t^{-1} dx$, $\theta^0= t^{-1} dt$.
These data describe~\cite{Mad00c} a noncommutative version of the
Lobachevski plane. The region around the line $t=1$ can be considered
as a vacuum. For the approximations of the previous section to be
valid we must rescale $t$ so that in a singular limit the vacuum
region becomes the entire space.  We can do this by setting
\begin{equation}
t = 1 + c t^\prime
\end{equation}
and consider the limit $c\to 0$. So that the geometry remain invariant
we must scale the metric. We do this by rescaling $\theta^0$
\begin{equation}
\theta^0 \mapsto c^{-1}\theta^0. 
\end{equation}
The commutation relations become then
\begin{equation}
[t^\prime,x] = c^{-1} h + h t^\prime
\end{equation}
and to leading order in $c$ the frame becomes
\begin{equation}
\theta^0 = (1-ct^\prime) dt^\prime, \qquad
\theta^1 = (1-ct^\prime) dx.
\end{equation}
From the definitions~(\ref{pert}) we
find that
\begin{equation}
\begin{array}{ll}
J_0^{01} = c^{-1} h, &
\epsilon I^{01} = ht^\prime,\\[6pt]
J_{0,01} = - c h^{-1}, &
\epsilon \Lambda^\alpha_\beta = c t^\prime \delta^\alpha_\beta
\end{array}
\end{equation}
and therefore we obtain the map $\sigma$ as defined in the previous
section. This example is not quite satisfactory since the cocycle
conditions~(\ref{dI}) are vacuous in dimension two.

\initiate 
\section{The WKB Ansatz}

We now suppose that the algebra $\c{A}$ is a tensor product 
\begin{equation}
\c{A} = \c{A}_0 \otimes \c{A}_\omega
\end{equation}
of a `slowly-varying' factor $\c{A}_0$ in which all amplitudes lie and
a `rapidly-varying' phase factor which is of order-of-magnitude
$\epsilon$ so that only functions linear in this factor can
appear. The generic element $f$ of the algebra is of the form then
\begin{equation}
f(x^\lambda, \phi) = f_0(x^\lambda) + 
\epsilon f_1(x^\lambda) e^{i\omega\phi}
\end{equation}
Because of the condition on $\epsilon$ these elements form an algebra.
We suppose that both $\Lambda$ and $I$ belong to $\c{A}_\omega$.  We
introduce the normal $\xi_\alpha = e_\alpha \phi$ to the surfaces of
constant phase.  From~(\ref{LI}) we find that
\begin{equation}
\h{\Lambda}^-_{\alpha\beta} =  - \h{I}_{\alpha\beta} + \xi_{[\alpha} A_{\beta]}
\end{equation}
The 2-form $\h{\Lambda}^-$ is, to within a constant, a plane-wave-type
solution to Maxwell's equations.  

The expression for the metric becomes
\begin{equation}
h^{\alpha\beta} = 
J^{-1}_{0\delta\gamma} I^{\gamma(\alpha}\eta^{\beta)\delta} +
\xi^{(\alpha} J_0^{\beta)}A^{\gamma}.                                 \label{h3}
\end{equation}
The Riemann tensor in the limit we
are considering, given by the expression
\begin{equation}
R_{\alpha\beta\gamma\delta} = \tfrac 14 \epsilon
\xi_{[\alpha} \ddot h_{\beta][\gamma}\xi_{\delta]},         \label{R2}
\end{equation}
depends only on the first term, linear in $I$.  We have defined
therefore a map
\begin{equation}
I^{\alpha\beta} \mapsto R^{\alpha}{}_{\beta\gamma\delta}
\end{equation}
from the algebra to the geometry.  Although there is a certain
amount of ambiguity in the definition of the map as far as the
components of the metric are concerned, this ambiguity drops from the
curvature. In Section~\ref{agm} we showed that all possible
perturbations of the differential calculi, except for the one which we
have chosen, leave the curvature invariant.

\initiate
\section{Dispersion relations}                   \label{dr}

The Ricci tensor for the perturbation $h_{\mu\nu}$ to a flat metric is
given by
\begin{equation}
R_{\alpha\beta} = - \tfrac 14 \epsilon\omega^2
(\xi^2 h_{\alpha\beta} - \xi^\gamma h_{\gamma(\beta} \xi_{\alpha)} +
h_\gamma^\gamma \xi_\alpha\xi_\beta).                 \label{R1aa}
\end{equation}
From Equation~(\ref{h3}) we see then that it is a linear expression in
the perturbation $I^{\mu\nu}$ to the commutation relations. 
The Einstein tensor for the perturbation is
given by
\begin{equation}
G_{\alpha\beta} = - \tfrac 14 \epsilon\omega^2 (\xi^2 \bar h_{\alpha\beta} + 
\bar h_{\gamma\delta} \xi^\gamma \xi^\delta g_{\alpha\beta} - 
\xi^\gamma \bar h_{\gamma(\beta} \xi_{\alpha)})         \label{E}
\end{equation}
in terms of 
\begin{equation}
\bar h_{\alpha\beta} = h_{\alpha\beta} - \tfrac 12 h g_{\alpha\beta}.
\end{equation}
From Equation~(\ref{h3}) we see then that both are linear expressions
in the perturbation $I^{\mu\nu}$ to the commutation relations.  The
vacuum field equations are given by
\begin{equation}
\xi^2 \bar h_{\alpha\beta} - 
\xi^\gamma \bar h_{\gamma(\alpha} \xi_{\beta)} +
\bar h_{\gamma\delta} \xi^\gamma \xi^\delta g_{\alpha\beta} = 0.\label{EVac}
\end{equation}
We require a plane-wave-like solution to the condition~(\ref{dI}), one
which is not the differential of a 1-form.  Wave-front surfaces are
2-surfaces and on such surfaces non-trivial 2-forms can exist. This is
however very formal since the surfaces in question are noncompact.
Within the context of the WKB approximation one can distinguish
between exact and non-exact closed 2-forms. If $\h{I}$ has frame
components
\begin{equation}
\h{I}_{\alpha\beta} = \h{I}_{0\alpha\beta} e^{i\omega\phi}
\end{equation}
then the differential has to leading order the components
\begin{equation}
(d\h{I})_{\alpha\beta\gamma} = i\omega (\xi_\alpha\h{I}_{\beta\gamma} +
\xi_\beta\h{I}_{\gamma\alpha} +\xi_\gamma\h{I}_{\alpha\beta}).   \label{dIabc}
\end{equation}
An example of a solution is
\begin{equation}
\h{I}_{\alpha\beta} = \tfrac 12 \epsilon_{\alpha\beta\gamma\delta}  
\h{I}^{*\gamma\delta}
\end{equation}
with 
\begin{equation}
\h{I}^{*\gamma\delta} \xi_\delta = 0.
\end{equation}
It follows from~(\ref{dIabc}) that if we multiply the cocycle
condition $d\h{I} = 0$ by $\xi_\alpha$ we obtain
\begin{equation}
\xi^2 \h{I}_{\alpha\beta} + 
\xi^\gamma \h{I}_{\gamma[\alpha}\xi_{\beta]} = 0.      \label{xidI}
\end{equation}
This equation is very similar in structure to~(\ref{EVac}) and contains the
essential information of the latter. From it one can read off the
dispersion relations. One sees that either $\h{I}$ is exact, that is
the metric perturbation is non-radiative, or $\xi^2 = 0$. We discuss
some of the details of this in the Appendix.

\initiate
\section{Nonlinearities}

Aided by a simplifying assumption, one can readily include the effects
of the higher-order terms neglected in the previous calculations. Let
$J^{\mu\nu}(z, \bar z)$ be an arbitrary antisymmetric matrix whose
elements belong to the subalgebra generated by two elements $z$ and
$\bar z$ of four $x^\mu$. Suppose further that $[z, \bar z] = 0$ so that the
subalgebra is abelian and to be explicit suppose that $z = x^3 + x^0$
and that $\bar z = x^3 - x^0$. To be consistent then we must suppose
further that $J^{03}=0$. Let $J^{\mu\alpha}$ be an arbitrary
invertible matrix of complex numbers. Reality conditions, which we
shall examine in more detail in a future publication force 
$J^{\mu\alpha}$ to be a real matrix to lowest order. We define the
geometry such that
\begin{equation}
[p_\alpha, x^\mu] = e^\mu_\alpha = J^{-1}_{\alpha\sigma} J^{\sigma\mu}.
\end{equation}
The notation is consistent since it follows that
\begin{equation}
 J^{\sigma\mu} = J^{\sigma\alpha} e^\mu_\alpha.
\end{equation}
Define now the commutators to be
\begin{equation}
[x^\mu, x^\nu] = i \kbar J^{\mu\nu} 
\end{equation}
This will be consistent provided the Jacobi identities
\begin{equation}
\epsilon_{\lambda\mu\nu\rho} 
J^{\lambda\alpha} e_\alpha J^{\mu\nu} = 0                \label{nlji}
\end{equation}
are satisfied. We shall return to this equation later. It follows
immediately that the Leibniz identities~(\ref{lr}) are satisfied.

Define finally
\begin{equation} 
i\kbar [p_\alpha, p_\beta] = - J^{-1}_{\alpha\mu} e^\mu_\beta.  
\end{equation}
The Lie algebra generated by the 8 elements is a consistent Lie
algebra provided the initial Jacobi identities are satisfied.  The
same logic as that which lead to the dispersion relation in
Section~\ref{dr} leads here to the conclusion that the matrix of
commutators must be a function only of $z$ (or $\bar z$) and that the
normal to the surface $z = z_0$ must be a null vector. Under Wick
rotation the matrix $J^{\mu\nu}$ would become a matrix of analytic
functions. 

\initiate
\section{Recapitulation}

In previous publications~\cite{Mad00c} we have shown that to a
noncommutative algebra defined by a commutator $J^{\mu\nu}$ and a
differential calculus defined by a frame $\theta^\alpha$ one can
associate (almost) a unique geometry defined by metric and connection.
The question of exactly what part of the information in the curvature
tensor comes from the commutator and which part from the frame remains
open. One might conjecture that if the couple
$(J^{\mu\nu},\theta^\alpha)$ defines one geometry with curvature map
$\mbox{Curv}(J^{\mu\nu},\theta^\alpha)$ and the couple
$(J^{\mu\nu},\theta^{\prime\alpha})$ a second geometry with curvature
map $\mbox{Curv}(J^{\mu\nu},\theta^{\prime\alpha})$ then one has
\begin{equation}
\mbox{Curv}(J^{\mu\nu},\theta^{\prime\alpha}) = 
\mbox{Curv}(J^{\mu\nu},\theta^\alpha).              \label{Curv}
\end{equation}
There are counter-examples to this conjecture; it is not true.

A weaker conjecture is that~(\ref{Curv}) is valid if the second frame
is a small perturbation
\begin{equation}
\theta^{\prime\alpha} = \theta^\alpha - 
\epsilon \Lambda^\alpha_\beta \theta^\beta
\end{equation}
of the first.  A yet weaker conjecture is that the
equality~(\ref{Curv}) is valid if the commutator is constant
$J^{\mu\nu}=J_0^{\mu\nu}$ and the initial frame is the exact frame
$\theta^\alpha = dx^\alpha$. We have shown that this is in fact true.
Furthermore we have shown that if the commutator is perturbed to
$J^{\mu\nu} = J_0^{\mu\nu} + \epsilon I^{\mu\nu}$ and the frame
$\theta^{\alpha}$ is consistently perturbed to $\theta^{\prime\alpha}$
then one has the equality
\begin{equation}
\mbox{Curv}(J^{\mu\nu},\theta^{\prime\alpha}) = 
\mbox{Curv}(I^{\mu\nu}).                            \label{Riem2}
\end{equation}
In other words the perturbation of the Riemann map depends only on the
perturbation $I$ of $J_0$ and not on its extension to the frame.
 
A second point which we have investigated is the status of the field
equations. In the `simplest' cases it would seem to be true that the
frame is dual to a set of derivations $e_\alpha$ of the algebra and
that these derivations are inner with associated momenta
$p_\alpha$. It would seem then that the theory contains only four
dynamical degrees of freedom. This is precisely the number of degrees
of freedom of the conformal tensor (in dimension four). One could
conjecture then that the Ricci tensor is fixed and calculable.
We have shown this to be the case if the 
algebra is a high-frequency perturbation of a flat background.

We have derived a relation between the structure of an
associated algebra as defined by the right-hand side $J$ of the
commutation relations between the generators $x^\mu$ on the one hand
and the metrics which the algebra can support, that is, which are
consistent with the structure of a differential calculus over the
algebra on the other. We have expressed this relation as the
map~(\ref{map}) from the frame to $J$ which defines the algebra. The
essential ingredients in the definition of the map are the Leibniz
rules and the assumption~(\ref{mod}) on the structure of the
differential calculus. Although there have been
found~\cite{FioMacMad01,CerFioMad00a,MacMadManZou03,BurMad05b}
numerous particular examples, there is not yet a systematic discussion
of either the range or kernel of the map. We have here to a certain
extent alleviated this, but only in the context of perturbation theory
around a vacuum and even then, only in the case of a high-frequency
wave. A somewhat similar relation has been found~\cite{Mad97c} in the case
of radiative, asymptotically-flat space-times.

\initiate
\section{Conclusion}

We started with a consistent flat-space solution to the constraints of
the algebra and of the geometry, a solution with the unusual property
that its momenta and position stand in a relation of simple duality, a
consequence of which is the fact that the Fourier-transformation is
local. We then perturbed both structures, the geometric and the
algebraic, in a seemingly arbitrary manner, but within the context of
linear-perturbation theory and requiring that the constraints remain
valid. We were able to completely solve the constraints of the
perturbation and exhibit a closed solution, which in the WKB
situation, implied that the Ricci tensor was necessarily flat.
However the seemingly general solutions we started with turned out,
all of them, to satisfy the simple duality of the original solution, a
fact which would tend to indicate that they were not really
sufficiently general. So whereas at best have presented a solid
indication that in the noncommutative context we have been persuing
the Ricci tensor can be considered as calculable; at least we have
indicated an interesting set of solutions to the algebro-geometic
problem which have the duality property of the original flat-space.

\initiate
\section{Appendix: WKB  cohomology}            

We briefly motivate here the notation used in
Section~\ref{dr}. We introduced the algebra of de~Rham forms
with a different differential inspired from the WKB approximation. The
differential can be introduced for all forms but we give the
construction only for the case of 2-forms. Let $f_{\alpha\beta}$ be a
2-form and define the differential $d_\xi$ of $f$ by the
Formula~(\ref{dI}). The interesting point is that the rank of the
cohomology module $H^2$, an elementary form of Spencer cohomology,
depends on the norm of $\xi$. Let $c$ be a 2-cocycle. Then
\begin{equation}
\xi_\alpha c_{\beta\gamma} +
\xi_\beta c_{\gamma\alpha} +\xi_\gamma c_{\alpha\beta} = 0.
\end{equation}
We multiply this by $\xi^\alpha$ to obtain the condition~(\ref{xidI}).
There are two possibilities. If $\xi^2 \neq 0$ then it follows
immediately that the 2-cocycle is exact. That is, $H^2 = 0$. If on the
other hand $\xi^2 = 0$ then there are cocycles which are not exact.
One can think of theses as plane-wave solutions to Maxwell's
equations. We can reformulate the result of Section~\ref{dr} as a
statement of the dependence of the Riemann tensor uniquely on the cohomology:
\begin{equation}
\mbox{Curv} = \mbox{Curv}[H^2].
\end{equation}

\initiate
\section*{Acknowledgment}

This article was started while JM was visiting the Humboldt
Universit\"at, Berlin and MB and he were visiting the A.~Einstein
Institut, Golm. They would like to thank D.~L\"ust as well as
H.~Nicolai and S.~Theisen for their hospitality during this period.
They would also like to thank J.~Mourad and S.~Waldmann for
enlightening comments.  The research was supported by the EPEAEK
programs ``Pythagoras'' and jointly funded by the European Union
(75\%) and the Hellenic state (25\%). The article assumed its present
form while JM was visiting the W.~Heisenberg Institut, M\"unchen. He
would like to thank F.~Meyer and J.~Wess for their hospitality during
this period.


\providecommand{\href}[2]{#2}\begingroup\raggedright\endgroup

\end{document}